\documentclass[aps,prl,floats,twocolumn,superscriptaddress,floatfix,showpacs]{revtex4}
\usepackage{epsf,graphicx}
\usepackage{amssymb}
\usepackage{amsmath}
\usepackage{eepic}
\usepackage[dvips]{color}

\begin{document}

\title{Possible nodeless $s^\pm$-wave superconductivity in twisted bilayer graphene}

\author{Zhe Liu}
\affiliation{Beijing National Laboratory for Condensed Matter Physics and
Institute of Physics, Chinese Academy of Sciences, Beijing 100190, China}
\author{Yu Li}
\affiliation{Beijing National Laboratory for Condensed Matter Physics and
Institute of Physics, Chinese Academy of Sciences, Beijing 100190, China}
\affiliation{University of Chinese Academy of Sciences, Beijing 100049, China}
\author{Yi-feng Yang}
\email[]{yifeng@iphy.ac.cn}
\affiliation{Beijing National Laboratory for Condensed Matter Physics and
Institute of Physics, Chinese Academy of Sciences, Beijing 100190, China}
\affiliation{University of Chinese Academy of Sciences, Beijing 100049, China}
\affiliation{Songshan Lake Materials Laboratory, Dongguan, Guangdong 523808, China}
\affiliation{Collaborative Innovation Center of Quantum Matter, Beijing 100190, China}

\date{\today}

\begin{abstract}
Recent discovery of superconductivity in the twisted bilayer graphene has stimulated numerous theoretical proposals concerning its exact gap symmetry. Among them, $d+id$ or $p+ip$-wave were believed to be the most plausible solutions. Here considering the superconductivity emerges near a correlated insulating state and may be induced by antiferromagnetic spin fluctuations, we apply the strong-coupling Eliashberg theory with both inter- and intraband quantum critical pairing interactions and discuss the possible gap symmetry in an effective low-energy four-orbital model. Our calculations reveal a nodeless $s^\pm$-wave as the most probable candidate for superconducting gap symmetry in the experimentally relevant parameter range. This solution is distinctly different from previous theoretical proposals. In particular, it contains interesting topological components in the valley space, which might be tuned by experimental manipulation of the valley degree of freedom.
\end{abstract}

\maketitle
\begin{figure}[t]
\centering\includegraphics[width=0.48\textwidth]{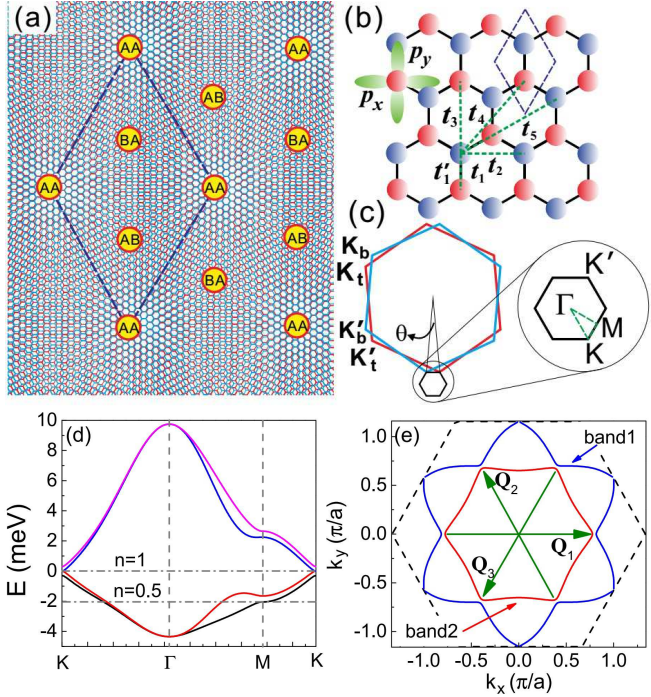}
\caption{(color online) (a) Illustration of the Moir\'e superlattice in TBLG with small twisted angle. (b) The honeycomb lattice for the tight-binding model used in this work. The $p_{x/y}$-like Wannier orbitals and the hopping parameters are labeled in the figure. In both (a) and (b), the regions enclosed by dashed lines represent a single super unit cell. (c) Illustration of the mini-Brillouin zone of TBLG. $K_t$ and $K_t'$ ($K_b$ and $K_b'$) represent the two inequivalent valleys of the top (bottom) graphene layer. $K$ and $K'$ represent the two inequivalent valleys of TBLG. $\Gamma$ and $M$ are the center and middle point of $K-K'$ line of the mini-Brillouin zone. (d) Effective low-energy band structures along the high symmetry lines in the mini-Brillouin zone. The dashed lines denote the Fermi energy for $n=0.5$ (half-filling) and 1, where $n$ is one fourth of the total occupation number. $n<1$ is called the hole-doping side. (e) Typical Fermi surfaces taken at $n=0.5$, where the two lower bands are half filled. $\mathbf{Q}_1$, $\mathbf{Q}_2$ and $\mathbf{Q}_3$ are the three nesting vectors used in our calculations.}
\label{fig1}
\end{figure}

Recent discovery of superconductivity in twisted bilayer graphene (TBLG) has attracted tremendous interest~\cite{Cao2018_1,Cao2018_2} in condensed matter community. Depending on the twisted angle, TBLG can exhibit a large variety of exotic phenomena~\cite{Henrard2007,Neto2007_2,Geim2008,Magaud2010,Andrei2010,Mele2010,MacDonald2010,Barticevic2010,MacDonald2011,Andrei2011,Mele2011,Koshino2012,Magaud2012,Neto2012,LHe2012,Beechem2012,Veuillen2012,Gonzalez2013,Oshiyama2014,Nori2015,LHe2015,Kaxiras2016,Cao2016,LHe2017,Koshino2017,MacDonald2017,San-Jose2017,Renard2018}. For small twisted angle, as shown in Fig.~\ref{fig1}(a), the lattice can form the so-called Moir\'{e} superlattice~\cite{Magaud2010,MacDonald2011,Beechem2012,Dean2018} with strongly renormalized low-energy Dirac fermions~\cite{Neto2007_2,MacDonald2011,Andrei2010,Mele2011,LHe2012,LHe2015,Kaxiras2016,Koshino2017,San-Jose2017}. At the so-called ``magic angles", their Fermi velocity can even be reduced to zero~\cite{Magaud2010,Andrei2010,MacDonald2011,LHe2012}. The superconducting phase in TBLG appears near the first magic angle ($\approx1.08^\circ$) on the hole-doping side ($n<1$), where the chemical potential is close to a van Hove singularity. The superconducting transition temperature is $T_c\approx 1.7\,\mathrm{K}$ with a renormalized electron bandwidth of about $10\,\mathrm{meV}$~\cite{Cao2018_1}. This was observed near a correlated insulating state at half-filling ($n=0.5$). It is believed that the superconductivity might be mediated by the associated antiferromagnetic quantum critical fluctuations~\cite{Cao2018_1,Cao2018_2,LFu2018_1,FYang2018}. Many theoretical efforts~\cite{CKXu2018,LFu2018_1,Senthil2018_1,Juricic,Scalettar2018,Baskaran,Phillips2018,CWang2018,TXMa,LZhang,Das,FYang2018,Heikkila,Spalek2018,Skryabin2018,Bascones,Senthil2,GMZhang,Mellado,Karrasch,LFu2018_2,LFu2018_3,Vishwanath1,CKXu2,Martin,Kuroki2018,Kaxiras,Walet,Senthil3,Stauber,SZLin2018,Bernevig,Betouras2018,Vafek_2018,CFang,Vishwanath2,XYGu2018,Nandkishore,Fernandes2018,RSHan,QHWang} have been devoted to understanding the effective low-energy models~\cite{LFu2018_1,Senthil2018_1,CWang2018,FYang2018,Mellado,LFu2018_3,CKXu2,Kuroki2018,Senthil3,Vafek_2018,Vishwanath2,Fernandes2018}, the nature of the insulating state~\cite{LFu2018_1,Senthil2018_1,Phillips2018,CWang2018,TXMa,FYang2018,Bascones,Karrasch,LFu2018_2,CKXu2,Kuroki2018,Stauber,Betouras2018}, and the gap symmetry of the superconducting phases~\cite{CKXu2018,Juricic,Scalettar2018,TXMa,LZhang,Das,FYang2018,Spalek2018,Mellado,Karrasch,Vishwanath1,SZLin2018,Betouras2018,Nandkishore,Fernandes2018,QHWang}, leading to the proposals of $d+id$-wave~\cite{CKXu2018,Scalettar2018,TXMa,LZhang,FYang2018,Spalek2018,Mellado,Karrasch,Nandkishore,Fernandes2018}, $p+ip$-wave~\cite{Juricic,Mellado,Stauber}, nodal $s$-wave~\cite{Das,Betouras2018}, $f$-wave~\cite{QHWang}, or even mixed gap symmetries~\cite{Vishwanath1,SZLin2018}. The nature of the superconducting phases is under heated debate.

The properties of the superconducting TBLG remind us some of the features of heavy fermion superconductors like $\mathrm{CeCu_2Si_2}$~\cite{Steglich1979,Steglich2011_1}. In CeCu$_2$Si$_2$, superconductivity is also mediated by magnetic quantum critical fluctuations, with $T_c\approx 0.6\,$K and a heavy electron band of the width of a few meV, similar to those of TBLG. For decades, superconductivity in $\mathrm{CeCu_2Si_2}$ has been believed to be $d$-wave~\cite{Steglich2011_1,Steglich2011_2,Steglich2011_3}, but recent refined experiments have revealed two nodeless gaps~\cite{Steglich2014,Steglich2016}. It was pointed out that such a gap structure may be resulted from strong interband quantum critical pairing interaction between coexisting electron and hole Fermi surfaces in CeCu$_2$Si$_2$~\cite{YuLi2018}. Since the low-energy effective model of TBLG may also have two bands~\cite{LFu2018_1,FYang2018}, we explore here the possibility of multiple superconducting gaps and study the detailed gap structures taking into account both intra- and interband pairing interactions. For this purpose, we adopt the strong-coupling Eliashberg approach and consider the magnetic quantum critical fluctuations as the candidate pairing glues~\cite{YuLi2018,Pines1990,Pines1991,Pines1992_2,Millis1992,Pines1992_1,Pines1994,Lonzarich2002,Golubov2009,Yang2014}. In comparison with the experiment, we find the most plausible gap symmetry to be nodeless $s^{\pm}$-wave in the observed range of superconductivity. Moreover, when projected into the valley basis, we obtain $f$-wave pairing in the valley-singlet component and  interesting topological features in the valley-triplet component.

Our low-energy effective model is constructed with one $p_x$-like and one $p_y$-like Wannier orbital on each site of the honeycomb Moir\'{e} superlattice \cite{LFu2018_1,LFu2018_3}, as schematically illustrated in Fig.~\ref{fig1}(b). The resulting four-orbital model describes the four low-energy bands separated by two gaps of about $50\,\mathrm{meV}$ from other bands~\cite{Cao2016}. At the first magic angle, the lattice constant is $a\approx134\,\mathrm{\AA}$. The mini-Brillouin zone of the superlattice is greatly reduced, as shown in Fig.~\ref{fig1}(c). Following Ref.~\cite{LFu2018_3}, we consider the hopping parameters up to the fifth nearest neighbors and take $t_1=2$, $t_2=0.3$, $t_3=0.05$, $t_4=0.15$, $t_5=0.12$, and $t_1'=0.1$ in units of meV, where $t_1$, $t_2$, $t_3$ and $t_4$ describe the hoppings between the same $p_x$ or $p_y$ orbitals, $t_5$ describes the hopping between $p_x$ and $p_y$ that lifts the degeneracy of the $K$ and $K'$ valleys, and $t_1'$ accounts for the hybridization between two valleys and generates a finite Dirac mass at $K$ and $K'$ points~\cite{LFu2018_1}. The parameters were chosen such that the bandwidth is of the order of $10\,\mathrm{meV}$ and the doping level for half-filling of the two lower bands ($n=0.5$) is close to the van Hove singularity. We have ignored the spin-orbit coupling (SOC) which is only $\sim 1\,\mu\mathrm{eV}$ in graphene and much smaller than the energy scale considered here~\cite{MacDonald2006,ZFang2007}. Figure~\ref{fig1}(d) plots the calculated band structures along high symmetry lines of the mini-Brillouin zone. The band structures at $K$ suggest the coexistence of massless and massive Dirac fermions, consistent with both symmetry analysis~\cite{LFu2018_1} and quantum oscillation experiment~\cite{Cao2018_1}. The Fermi surfaces are nested near the van Hove singularity, as depicted in Fig.~\ref{fig1}(e) with the nesting wave vectors $\mathbf{Q}_1=4\pi/3a(1,0)$, $\mathbf{Q}_2=4\pi/3a(-1/2,\sqrt{3}/2)$ and $\mathbf{Q}_3=4\pi/3a(-1/2,-\sqrt{3}/2)$. A Lifshitz transition occurs when the chemical potential is tuned across the van Hove point.

To study the pairing symmetry of superconductivity, we apply the strong-coupling Eliashberg theory and consider two separated Fermi surfaces. The interband (finite momentum) pairing was ignored due to the absence of Fermi surface overlap~\cite{YuLi2018}. Near $T_c$, the linearized Eliashberg equations can be written as
\begin{equation}
\begin{split}\label{eliash1}
\lambda\phi_\mu(\mathbf{k},i\omega_n)= & -\pi T\sum_{\nu,m}\oint_{\mathrm{FS}_\nu}dk'_{\parallel}\phi_{\nu}(\mathbf{k}',i\omega_m)\\
& \times\frac{V^{\mu\nu}(\mathbf{k}-\mathbf{k}',i\omega_n-i\omega_m)}{(2\pi)^2v_{\mathbf{k}_{\mathrm{F}}'}|\omega_{m}Z_{\nu}(\mathbf{k}',i\omega_{m})|},
\end{split}
\end{equation}
with
\begin{equation}
\begin{split}
Z_\mu(\mathbf{k},i\omega_n)= & 1+\frac{\pi T}{\omega_n}\sum_{\nu,m}\oint_{\mathrm{FS}_\nu}dk'_\parallel\mathrm{sgn}(\omega_{m})\\
& \times\frac{V^{\mu\nu}(\mathbf{k}-\mathbf{k}',i\omega_n-i\omega_{m})}{(2\pi)^2v_{\mathbf{k}_{\mathrm{F}}'}},
\end{split}
\end{equation}
where $\mu$ and $\nu$ denote the band indices, $\omega_n$ is the fermionic Matsubara frequency, $Z_{\mu}(\mathbf{k},i\omega_n)$ is the renormalization function, and $\phi_\mu(\mathbf{k},i\omega_n)$ is related to the gap function through $\phi_\mu(\mathbf{k},i\omega_n)=Z_\mu(\mathbf{k},i\omega_n)\Delta_{\mu}(\mathbf{k},i\omega_n)$. $V^{\mu\nu}(\mathbf{k}-\mathbf{k}',i\omega_n-i\omega_m)$ represents the intra- or interband scattering of the Cooper pairs. This is an eigenvalue equation and each of its eigen solutions corresponds to a candidate pairing channel. The superconducting symmetry is determined by the leading channel with the largest eigenvalue $\lambda$ at $T_c$. For quantum critical superconductivity, the pairing interactions take the phenomenological form \cite{Pines1990,YuLi2018},
\begin{equation}
V^{\mu\nu}(\mathbf{q},i\nu_n)=\frac{V_0^{\mu\nu}}{1+\xi^2(\mathbf{q}-\mathbf{Q})^2+|\nu_n|/\omega_{sf}},
\end{equation}
where $\xi$ is the correlation length, $\mathbf{Q}$ is the antiferromagnetic wave vector and $\omega_{sf}$ is the characteristic energy of the  quantum critical fluctuations. The values of $\xi$ and $\omega_{sf}$ were chosen such that the effective magnetic Fermi energy, $\Gamma_{sf}=\omega_{sf}(\xi/a)^2=4.3\,\mathrm{meV}$, is approximately equal to the bandwidth of the two lower bands. $V_0^{\mu\nu}$ are free parameters to be determined experimentally. The resulting gap structures only depend on the relative strengths, $r_{11}=V^{11}_0/V^{22}_0$ and $r_{12}=V^{12}_0/V^{22}_0$, while their absolute values play no essential role.

\begin{figure}[t]
\centering\includegraphics[width=0.48\textwidth]{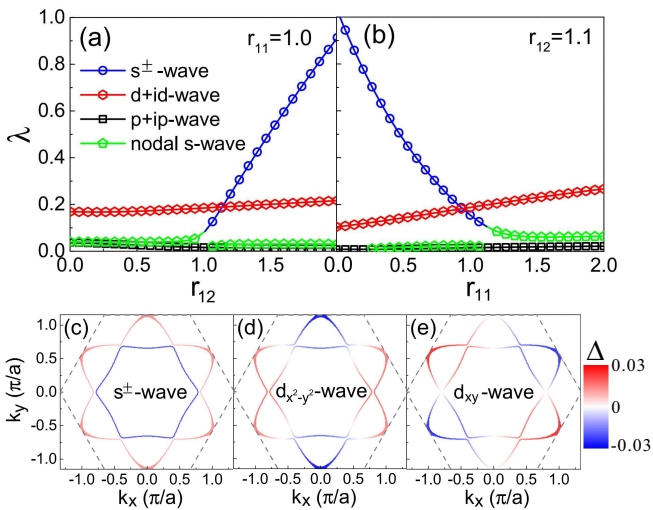}
\caption{(color online) Evolution of the eigenvalues $\lambda$ of four leading solutions with (a) $r_{12}$ for fixed $r_{11}=1.0$ and (b) $r_{11}$ for fixed $r_{12}=1.1$. (c)-(e) plot the gap distribution on the Fermi surfaces for a typical nodeless $s^\pm$-wave solution and the degenerate $d_{x^2-y^2}$ and $d_{xy}$-wave solutions, respectively. We take as an example the results at $n=0.5$. }
\label{fig2}
\end{figure}

To solve the Eliashberg equations, we further approximate $\Delta_\mu(\mathbf{k},i\omega_n)=\Delta_\mu(\mathbf{k},i\pi T_c)$ and $Z_\mu(\mathbf{k},i\omega_n)=Z_\mu(\mathbf{k},i\pi T_c)$ and use 2048 Matsubara frequencies and $240\times240$ $\mathbf{k}$-meshes in the mini-Brillouin zone. Typical solutions of the eigenvalue equations are plotted in Fig.~\ref{fig2}. The leading solution is a spin-singlet state with nodeless $s^\pm$ or doubly degenerate $d$-wave gap symmetry. This may be understood by recalling that the superlattice has a $D_3$ symmetry group with two one-dimensional and one two-dimensional irreducible representations. The $s^\pm$-wave and degenerate $d$-wave solutions correspond to the $A_1$: 1, $k_x^2+k_y^2$, ... and $E: (k_x,\ k_y)$, $(k_x^2-k_y^2,\ k_xk_y)$ representations, respectively. A degenerate $p$-wave solution is also allowed in the $E$ representation but never becomes dominant in the relevant parameter range. Figure~\ref{fig2}(a) shows the four leading values of $\lambda$ with varying $r_{12}$ at fixed $r_{11}=1.0$. For small $r_{12}$, the $d$-wave solutions dominate, indicating that the intraband interaction favors $d$-wave superconductivity on both Fermi surfaces. Since $d_{xy}$ and $d_{x^2-y^2}$ are degenerate, the true gap function here is their linear combination, yielding a chiral $d+id$-wave pairing symmetry~\cite{FYang2018,Karrasch}. While for large $r_{12}\gtrsim1.2$, the leading solution is nodeless $s^\pm$-wave, induced by a relatively strong interband interaction, as is in pnictide or some heavy fermion superconductors~\cite{Machida2016,YuLi2018,Stewart2017}. A nodal $s$-wave solution never wins out in the whole parameter range. Figure~\ref{fig2}(b) plots the leading eigenvalues at fixed $r_{12}=1.1$. With increasing $r_{11}$, the dominant $d$ and nodeless $s^\pm$-wave solutions exhibit opposite tendencies, reflecting their different physical origins owing to the intra- and interband quantum critical pairing interactions, respectively. The momentum distribution of the two solutions are plotted in Figs.~\ref{fig2}(c)-(e). Although the $d_{xy}$ and $d_{x^2-y^2}$ components have nodes, the combined $d+id$ solution is fully gapped. Therefore, both leading solutions are nodeless and may be hard to distinguish in the scanning tunneling experiment.

\begin{figure}[t]
\centering\includegraphics[width=0.46\textwidth]{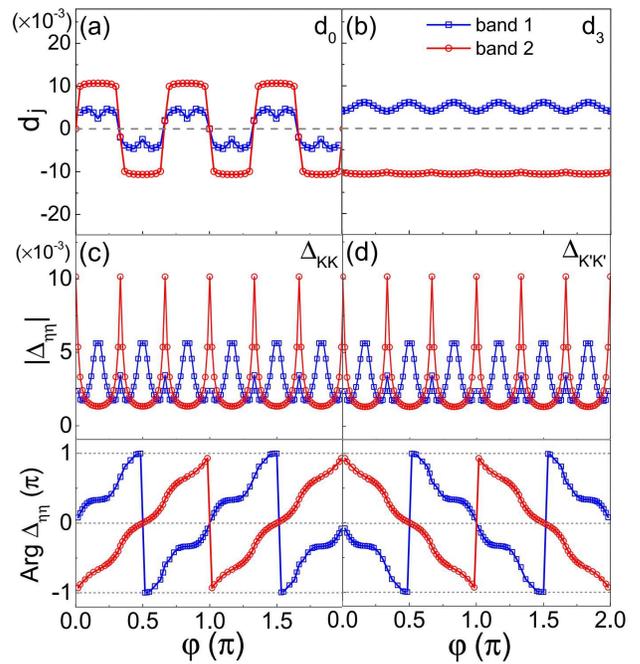}
\caption{(color online) Variation of the gap functions in the valley space with the azimuthal angle $\varphi$, showing (a) the valley-singlet component $d_0$, (b) the valley-triplet component $d_3$, (c) $\Delta_{KK}=-d_1+id_2$, and (d) $\Delta_{K'K'}=d_1+id_2$ for a typical nodeless $s^\pm$-wave solution of the band basis. The upper and lower panels of (c) and (d) plot the magnitude and argument of the two intravalley components, respectively.}
\label{fig3}
\end{figure}

While these solutions look simple, they contain certain topological features in the valley space. This may be analyzed considering $\Delta_{\eta\eta'}(\mathbf{k})=\sum_{\mu}a_{\eta\mu}(\mathbf{k})a_{\eta'\mu}(-\mathbf{k})\Delta_\mu(\mathbf{k})$, under a unitary transformation from band ($\mu$) to valley ($\eta$) bases, $c_{\eta\sigma}(\mathbf{k})=\sum_\mu a_{\eta\mu}(\mathbf{k})c_{\mu\sigma}(\mathbf{k})$. The resulting gap function can be quite generally decomposed into the valley-singlet and triplet components through $\Delta_{\eta\eta'}(\mathbf{k})=\sum_j[d_j(\mathbf{k})\tau_j i\tau_2]_{\eta\eta'}$, where $\tau_j$ are the unit matrix for $j=0$ and the Pauli matrices for $j=1,\, 2,\, 3$. The magnitudes of $d_0$ and $\mathbf{d}=(d_1, d_2, d_3)$ represent the relative importance of the valley-singlet and triplet contributions. Figure~\ref{fig3} plots the distribution of $d_j(\mathbf{k})$ on the Fermi surfaces for a typical solution of nodeless $s^\pm$-wave as a function of the azimuthal angle $\varphi$. For spin-singlet pairing, the valley-singlet (triplet) requires odd (even) gap symmetry in the momentum space due to the Pauli principle. As shown in Fig.~\ref{fig3}(a), $d_0$ indeed is an odd function in the momentum space. In particular, it has the same sign on both Fermi surfaces at the same azimuthal angle but the overall $\varphi$-dependence exhibits an $f$-wave manner. Thus the nodeless $s^\pm$-wave solution in the band basis contains a valley-singlet $f$-wave component. The angle dependence of the valley-triplet component is analyzed in Figs.~\ref{fig3}(b)-(d). Quite unexpectedly, while $d_3(\mathbf{k})$ is real and varies only slightly with momentum on both Fermi surfaces, $d_1(\mathbf{k})$ and $d_2(\mathbf{k})$ that correspond to intravalley pairing exhibit unusual topological characters. To see this, we introduce the gap functions on each valley individually, $\Delta_{KK}=-d_1+id_2$ and $\Delta_{K'K'}=d_1+id_2$, and plot the angle dependence of their amplitudes and phases in Figs.~\ref{fig3}(c) and \ref{fig3}(d). A phase change of $4\pi$ or $-4\pi$ is revealed as $\varphi$ varies from $0$ to $2\pi$, which is a characteristic feature of topological superconductor with time reversal symmetry \cite{Ishida2012}. We thus conclude that the valley-triplet component has $d\pm id$-wave symmetry in the momentum space. We should note that the existence of $\Delta_{KK}$ and $\Delta_{K'K'}$ is associated with the presumption of valley hybridization in our model Hamiltonian \cite{LFu2018_1}. The relative importance of different valley components may be tuned by experimental manipulation of the valley degree of freedom~\cite{QNiu2007,Beenakker2007}. Absent valley hybridization, the gap function becomes topologically trivial. Similar analysis may be applied to the $d+id$-wave solution in the band basis (not shown), where the $d_3$ and $d_1+id_2$ components exhibit the phase change of $-4\pi$ and $-8\pi$, respectively, implying its topological nature as chiral superconductivity~\cite{Ishida2012}.

\begin{figure}[t]
\centering\includegraphics[width=0.48\textwidth]{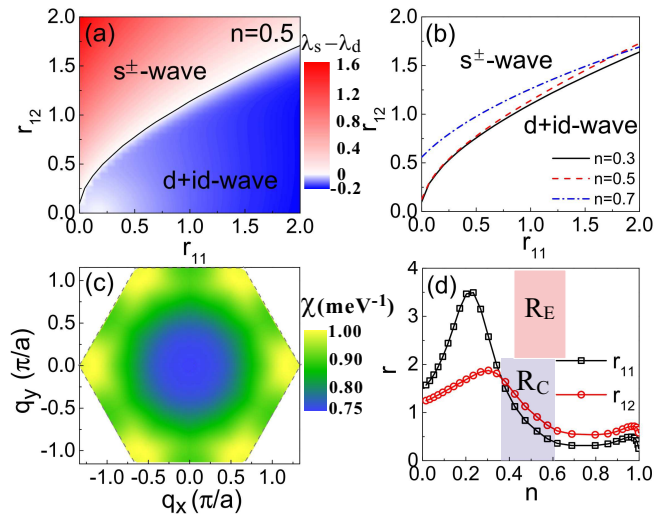}
\caption{(color online) (a) A typical theoretical phase diagram taken with $n=0.5$. The color is calibrated by the difference between the eigenvalues of the nodeless $s^\pm$ and $d+id$-wave solutions. The solid line marks the boundary of the two phases. (b) Variation of the phase boundary for different doping levels, showing only slight change with $n$. (c) The total spin susceptibility $\chi(\mathbf{q})$ evaluated from the Lindhard function, showing maxima at nesting wave vectors. (d) Variation of the estimated $r_{11}$ and $r_{12}$ as a function of the doping level. The shaded area $\mathrm{R}_\mathrm{C}$ denotes the range where the nodeless $s^\pm$-wave becomes dominant, while the shaded area $\text{R}_\text{E}$ marks the region of superconductivity observed in experiment. The insulating phase (not indicated) is a very narrow region near half-filling within $\mathrm{R}_\mathrm{E}$.}
\label{fig4}
\end{figure}

Our results are summarized in Fig.~\ref{fig4}(a) on a generic phase diagram with $r_{11}$ and $r_{12}$ as tuning parameters. There are two  regions governed by the nodeless $s^\pm$-wave for large $r_{12}$ and the $d+id$-wave for small $r_{12}$. The overall phase boundary is only slightly shifted for different doping levels as shown in Fig.~\ref{fig4}(b). Although an exact estimate of the relative importance of the intra- and interband quantum critical  pairing interactions is not possible at this stage, some preliminary argument might still be made by considering $V^{\mu\nu}_0 \propto \text{Re}\chi^{\mu\nu}(\mathbf{Q})$, where $\chi^{\mu\nu}(\mathbf{Q})$ is the static spin susceptibility at the ordering wave vector $\mathbf{Q}$ and may be estimated under the first order approximation using the Lindhard function,
\begin{equation}
\chi^{\mu\nu}(\mathbf{q})=\sum_{\mathbf{k}}\frac{f_{FD}(\varepsilon_{\mu\mathbf{k}})-f_{FD}(\varepsilon_{\nu,\mathbf{k}+\mathbf{q}})}{\varepsilon_{\nu,\mathbf{k}+\mathbf{q}}-\varepsilon_{\mu\mathbf{k}}+i\delta},
\end{equation}
where $f_{FD}$ is the Fermi-Dirac distribution function and $\varepsilon_{\mu\mathbf{k}}$ is the dispersion of the $\mu$-th band. Taking once again $n=0.5$ as an example, we plot in Fig.~\ref{fig4}(c) the real part of $\chi(\mathbf{q})=\sum_{\mu\nu}\chi^{\mu\nu}(\mathbf{q})$ as a function of $\mathbf{q}$. As expected, $\text{Re}\chi(\mathbf{q})$ reaches maximum when $\mathbf{q}$ approaches the nesting wave vectors ($\mathbf{Q}_1$, $\mathbf{Q}_2$, $\mathbf{Q}_3$) at the corners of the mini-Brillouin zone. The corresponding $r_{11}$ and $r_{12}$ can also be estimated and plotted in Fig.~\ref{fig4}(d) for generic $n$. A direct comparison with the phase diagram in Fig.~\ref{fig4}(b) gives the lower shaded area $\mathrm{R}_\mathrm{C}$ ($0.4\lesssim n\lesssim0.6$) in which the nodeless $s^\pm$-wave solution dominates. Experimentally, superconductivity was observed in the upper shaded area labeled as $\mathrm{R}_\mathrm{E}$. Their overlap implies that nodeless $s^\pm$-wave is the most plausible gap symmetry for the superconducting TBLG near half-filling on the hole-doping side.

Our prediction of the nodeless $s^\pm$-wave pairing symmetry is deduced from an effective low-energy four-band model with quantum critical pairing interactions and two coexisting Fermi surfaces due to valley hybridization. In previous studies \cite{CKXu2018,Scalettar2018,TXMa,LZhang,FYang2018,Spalek2018,Mellado,Karrasch,Nandkishore,Fernandes2018}, a $d+id$-wave solution was often obtained without considering the presence of strong interband interaction. Absent valley hybridization, $p+ip$ \cite{Juricic,Mellado,Stauber} or nodal $s$-wave solutions \cite{Das,Betouras2018} have also been proposed depending on the topology of the Fermi surfaces. On the other hand, a more sophisticated analysis using functional renormalization group (fRG) was shown to yield an $f$-wave solution \cite{QHWang}. Actually, this latter work might be consistent with the valley-singlet component in our nodeless $s^\pm$-wave solution, as valley hybridization was neglected in the fRG calculations. Further experiments are needed to examine these various possibilities. However, we should note that both the nodeless $s^\pm$ and $d+id$-waves are nodeless in the momentum space. Therefore, it might be difficult to distinguish them using the usual scanning tunneling or angle-resolved photoemission spectroscopies. In this respect, phase sensitive measurements using Josephson devices for example \cite{Harlingen1995} have recently been applied to the investigation of 2D topological superconductivity \cite{Lombardi2017}, and Kerr rotation experiments may also be used to detect time reversal symmetry breaking in $d+id$ or $p+ip$-pairings~\cite{Kapitulnik2015}.

To summarize, we explore possible gap symmetry of the superconductivity observed recently in TBLG based on a four-orbital model using the strong-coupling Eliashberg equations with a quantum critical form of the intra- and interband pairing interactions. We find a leading nodeless $s^\pm$-wave solution for strong interband interaction and a dominant $d+id$-wave solution originating primarily from the intraband pairing interaction. Both exhibit interesting topological characters in the valley basis. A direct comparison between our theoretical and experimental parameter ranges suggests that the nodeless $s^\pm$-wave is the most plausible candidate for the pairing symmetry in superconducting TBLG, different from previous theoretical proposals.

This work was supported by the National Key R\&D Program of China (Grant No. 2017YFA0303103), the National Natural Science Foundation of China (NSFC Grant Nos. 11774401, 11522435), the State Key Development Program for Basic Research of China (Grant No. 2015CB921303), the Strategic Priority Research Program (B) of the Chinese Academy of Sciences (Grant No. XDB07020200) and the Youth Innovation Promotion Association of CAS.

\end{document}